\documentstyle[psfig,aps,prl,epsf]{revtex}
\begin{document}
\twocolumn[\hsize\textwidth\columnwidth\hsize\csname %
@twocolumnfalse\endcsname
\title{Variational study of the extended Hubbard-Holstein model
on clusters of variable site spacing}
\author{Marcello Acquarone$^a$, Mario Cuoco$^b$, Canio Noce$^b$ and Alfonso Romano$^b$}
\address{$^a$CNR-GNSM, Unit\`a I.N.F.M. di Parma, Dipartimento di
Fisica dell'Universit\`a di Parma, I-43100 Parma, Italy}
\address{$^b$Unit\`a I.N.F.M. di Salerno, Dipartimento di Scienze Fisiche
''E.R. Caianiello'',\\ Universit\`a di Salerno, I-84081 Baronissi
(Salerno), Italy}
\date{\today}
\maketitle
\begin{abstract}
We study the complete extended Hubbard-Holstein Hamiltonian on a
four-site chain with equally spaced sites, with spacing-dependent
electronic interaction parameters evaluated in terms of Wannier
functions built from Gaussian atomic orbitals.  By successive
application of generalized displacement and squeezing
transformations, an effective polaronic Hamiltonian is obtained,
containing a purely phonon-induced, long-range intersite
charge interaction. The phase diagrams for $N=1,2,3,4$ electrons
have been determined by variational minimization of the sum of the
electronic and the phononic energy, paying special attention to
the effects of the above mentioned phonon-induced long-range
interaction. To characterize the physics of each ground state, we
have evaluated how variations of the site spacing affect the
behavior of several correlation functions of experimental
interest, with the aim of representing lattice deformation
effects, either spontaneous or induced by external pressure.

\end{abstract}
 ]

\narrowtext
\section{Introduction.}
The model describing correlated electrons in a narrow band
interacting locally with lattice deformations, the so-called
Holstein-Hubbard model, has been studied intensively in the last
years both on infinite lattices\cite{hollit}--\cite{SQ} and on
small clusters\cite{clusters}--\cite{chunli}. The classes of
materials to which the model is assumed to apply, such as, for
instance, the high-temperature superconducting cuprates or the
manganites showing colossal magnetoresistance, are characterized
by rather strong electron-phonon coupling, so that a perturbative
treatment is in general unreliable. In that case, one can apply to
the interacting electron-phonon Hamiltonian a unitary
``displacement'' transformation\cite{wagner} generated by a
suitably chosen anti-hermitian operator $D$. From the displaced
Hamiltonian $\tilde{H}=e^D \, H \, e^{-D}$ a non-perturbative
effective Hamiltonian for phonon-dressed electrons (polarons) can
be obtained by averaging $\tilde{H}$ over an appropriate phonon
wavefunction. In recent years, the use of variationally optimized
"squeezed" phonon states of the form $e^{-S}|0_{ph}\rangle$, with
$|0_{ph}\rangle$ being the phonon vacuum, has proved to be
advantageous \cite{trapper,SQ,fehske,weimin} over methods using
simple harmonic oscillator states. However, as the operators $D$
and $S$ on a lattice with $L$ sites each contain $L$ undetermined
$c$-numbers, one for each phonon wavevector ${q}$, such procedure
requires to set all such $2L$ displacement and squeezing
parameters to their optimal values. A variational determination on
an infinite lattice is therefore impossible, and one has to adopt
perturbative techniques\cite{localdelta,wang}. For this reason,
the most popular approach has been to neglect the $q$-dependence
of the displacement parameters, by assuming $\delta_q=1$ for any
$q$. This is the standard Lang-Firsov approximation (LFA)
\cite{lf}, which, however, has at least two limitations. On one
side, it neglects any inter-site phonon correlation, preventing
the study of how the charge on a given site can induce
deformations also on different sites, possibly giving rise to
extended polarons\cite{emin}. On the other side, one can
show\cite{mazako} that, for non-dispersive phonons, the LFA leads
to the vanishing of a purely phonon-induced intersite charge
interaction (the ``residual interaction'' first introduced in
Ref.\onlinecite{alexran}), in principle present in $\tilde{H}$.
One concludes that the physics of the system is not faithfully
represented in LFA. If, instead of infinite lattices, one
considers small clusters\cite{clusters}, the numerical complexity
is reduced, and one can rigorously take into account the
$q$-dependence of all the variational parameters. In this way the
limitations of LFA can be avoided, and, in particular, the
inter-site correlations between charges and deformations, as well
as the effects of the purely phonon-induced interaction terms, can
be satisfactorily taken into account.

This is actually one of the points on which the present work
focuses. Indeed, we shall first derive the effective polaronic
Hamiltonian from a bare Hamiltonian which  includes in the
electronic part all the one- and two-body interactions compatible
with a non-degenerate band of electrons, and has an
electron-phonon coupling of the Holstein type. As far as the set
of electronic interactions is concerned, the present work extends
previous studies, where the only two-body electronic coupling
terms included in the bare Hamiltonian were the on-site and
inter-site Coulomb charge interactions. Then we shall solve
exactly the effective model on a four-site chain with constant
intersite distance $a$ and closed boundary conditions. Considering
that in this case the wavevector $q$ takes four values, given by
$-\pi/2a$, 0, $\pi/2a$ and $\pi/a$, the independent displacement
and squeezing parameters are six in total, as
$\delta_{\pi/2a}=\delta_{-\pi/2a}$ and
$\alpha_{\pi/2a}=\alpha_{-\pi/2a}$. They will be simultaneously
and independently determined by exact diagonalization of the
effective Hamiltonian, followed by variational optimization of the
ground state energy, for given, but arbitrary, site spacing and
given shape of the Wannier functions, as discussed below. This
will allow us, on the one hand, to discuss the interplay between
charges and phonons both on the same site and on different sites,
and, on the other hand, to properly take into account the effect
of the phonon-induced inter-site charge interaction.

Another novel aspect of the present work is that, as we model the
dependence on the site spacing of all the electronic interaction
parameters, we can study the phase diagram and the correlation
functions at any given filling as functions of the intersite
distance. Our results can therefore give qualitative indications
relevant both to measurements under mechanical or chemical
pressure, and to materials showing ``stripes''\cite{stripemat} or
other forms of spontaneously coupled charge and structural
inhomogeneities\cite{bozin}, where regions with different lattice
parameters, and different electronic character, have been reported
to coexist.

The paper is organized as follows. In Section 2, we define the
model and derive the effective polaronic Hamiltonian. The results
of the numerical analysis on a four-site closed chain for
different fillings $N=1,2,3,4$ are discussed in Section 3. Section
4 is devoted to the conclusions.

\section{The model and the effective Hamiltonian.}

Our model Hamiltonian  $H = H_{el}+H_{el-ph}+H_{ph}$ consists of a
generalized Hubbard Hamiltonian for correlated electrons coupled
to phonons of frequency $\Omega_q$. The bare electronic
Hamiltonian reads
\begin{eqnarray}
H_{el} & = & \sum_{j\sigma }\epsilon _{j}^{~} n_{j\sigma}^{~}
-\sum_{jl\sigma}[t_{jl}^{}-X^{}_{jl}(n_{j,-\sigma}^{~}+n_{l,-\sigma}^{~})]
c_{j\sigma }^{\dagger }c_{l\sigma}^{~} \nonumber \\ && +\,
U\sum_{j}n_{j\uparrow}^{~}n_{j\downarrow }^{~} + \sum_{jl}V_{jl}
n_{j}^{~}n_{l}^{~}+ \sum_{jl}J_{jl}\,{\bf S}_{j}^{~} \cdot {\bf
S}_{l}^{~} \nonumber \\ && +\sum_{jl}P_{jl}
\left(c_{j\uparrow}^{\dagger }c_{j\downarrow}^{\dagger}
c_{l\downarrow}^{~} c^{~}_{l\uparrow}+{\it H.c.} \right)
\label{Hamel}
\end{eqnarray}
where $c_{j\sigma}^{\dag}$ ($c_{j\sigma}$) creates (annihilates)
an electron with spin $\sigma$ on site $j$, and ${\bf S}_j$ and
$n_j$ are the total spin and charge on site $j$, respectively.
Following the procedure reported in Ref.\onlinecite{acquarone},
the bare electronic interactions $\epsilon,t,U,V,J,X$, are
parametrized by means of Wannier function built from Gaussian
functions
$\phi_j(r)=(2\Gamma^2/\pi)^{3/4}\exp[-\Gamma^2(r-R_j)^2]$
modelling the atomic orbitals on the $j$-th site in the position
$R_j$. The Wannier functions, and by consequence the interactions,
depend on the intersite distance $a$ and the width $\Gamma^{-1}$
of the orbitals. The explicit expressions of
$\epsilon,t,U,V,J,P,X$ that we use in this paper have been
evaluated in the case of a dimer and are given in
Ref.\onlinecite{acquarone}. This allows us to estimate all the
electronic parameters in an ab-initio type of scheme.

The interaction $H_{el-ph}$ between the electrons and the phonons
is assumed of the Holstein type $H_{el-ph}= G\sum_{j\sigma
}u_{j}^{ }n_{j\sigma}$, where £$u_j$ is the lattice deformation at
site $j$. By introducing the characteristic length
${\cal{L}}_q=\sqrt{\hbar/2M\Omega_q}$ and the renormalized
electron-phonon coupling $g_q\equiv G{{\cal{L}}}_q$, in second
quantization $H_{el-ph}$ takes the form
\begin{equation}
H_{el-ph}=\sum_q \, g_q\,(b^{\dagger}_{-q}+b^{ }_{q})\,n_q^{~}
\end{equation}
where $n_q = L^{-1/2}\sum_{j,\sigma} n_{j,\sigma}^{~}\exp[iqR_j]$,
$L$ being the number of lattice sites. Finally, the free phonon
term
\begin{equation}
H_{ph}=\sum_q\hbar\Omega_q(b^{\dagger}_{q}b^{ }_{q}+1/2)
\end{equation}
completes the model Hamiltonian.

The general procedure to obtain the effective polaronic
Hamiltonian and to identify its ground state consists in the
following steps \cite{acquarone}:

(i) a unitary transformation generated by a suitably chosen
anti-hermitian operator $D$ is used to map the model Hamiltonian
$H$ onto the "displaced" Hamiltonian $\tilde{H}=e^DHe^{-D}$
\cite{wagner}. We choose
\begin{equation}
D= \sum_{q}\,\delta_q\,
{{g_q}\over{\hbar\Omega_q}}\,(b^+_{-q}-b_q)\,n_q
\end{equation}
where the parameters $\delta_{q}$, defined on the $L$ wavevectors
of the first Brillouin zone, measure the degree of ``displacement"
of the corresponding phonon mode. Their real-space expressions
$\delta_{jl}$, associated with the coupling between charge on site
$j$ and phonons on site $l$, are given by
$\delta_{jl}=(1/L)\sum_{q}\delta_{q}\exp[iq(R_j-R_l)]$. It is
evident that assuming $\delta_{q}=1$ for any $q$ (as it is done in
the LFA) implies $\delta_{jl}=0$ if $R_j\ne R_l$, forcing the
effective interactions to be strictly local;

(ii) by averaging $\tilde{H}$ over the squeezed phonon state
$|\Psi_{ph}\rangle$ \cite{weimin,zheng} the phononic degrees of
freedom are eliminated, yielding the effective polaronic
Hamiltonian. The squeezed state is defined as
$|\Psi_{ph}\rangle=e^{-S}|0_{ph}\rangle$
\cite{fehske,weimin,zheng}, where $|0_{ph}\rangle$ is the harmonic
oscillator vacuum state, such that $b_q|0_{ph}\rangle=0$, and  the
squeezing generator $S$, expressed in terms of the $L$ variational
parameters $\alpha_q$ (or, in real space,
$\alpha_{jl}=(1/L)\sum_q\alpha_q\exp[iq(R_j-R_l)]$) reads:
\begin{equation}
S=\sum_q \,\alpha_q \left( b^\dagger_qb^\dagger_{-q}-b_q b_{-q}
\right) =\sum_{j,l}\alpha_{jl} \left(b^\dagger_j b^\dagger_{l}-b_j
b_{l} \right)
\end{equation}
This procedure leads to an effective Hamiltonian $H^\ast$
describing phonon-dressed electrons (which we shall generally
designate as polarons) having the form
\begin{eqnarray}
H^\ast & = & \sum_{q} \hbar\Omega _{q}\left[ \sinh ^{2}\left(
\alpha_{p}\right) +\frac{1}{2}\right]+ \sum_{j\sigma }\epsilon
_{j}^{\ast }\,n_{j\sigma} \nonumber \\ & &
-\sum_{jl\sigma}[t_{jl}^{\ast}-X^\ast_{jl}
(n_{j,-\sigma}+n_{l,-\sigma})]\, c_{j\sigma }^{\dagger
}c_{l\sigma} \nonumber \\ & & + U^{\ast}\,\sum_{j}n_{j\uparrow
}n_{j\downarrow } + \sum_{jl}V_{jl}^{\ast}\, n_{j}n_{l}+
\sum_{jl}J^\ast_{jl}\,{\bf S}_{j}\cdot{\bf S}_{l} \nonumber \\ & &
+\sum_{jl}P_{jl}^\ast\left(c_{j\uparrow}^{\dagger
}c_{j\downarrow}^{\dagger} c_{l\downarrow}c_{l\uparrow}+{\it H.c.}
\right) \label{effHamphonpol}
\end{eqnarray}
with the renormalized interactions
\begin{equation}
\epsilon _{j}^{\ast }=\epsilon _{j}-\frac{1}{L}\sum_{q}
{{g^2_q}\over{\hbar\Omega_{q}}}\, \delta _{q}\left( 2-\delta
_{q}\right) \label{epsiloneff}
\end{equation}
\begin{equation}
U_{j}^{\ast }=U_{j}-2\left( \frac{1}{L}\right) \sum_{q}
{{g^2_q}\over{\hbar\Omega_{q}}}\, \delta _{q}\left( 2-\delta
_{q}\right) \label{ueffective}
\end{equation}
\begin{equation}
V_{jl}^{\ast }=V_{jl}-\frac{1}{L}\sum_{q} {{g^2_q}\over{
\hbar\Omega_{q}}}\, \delta _{q}\left( 2-\delta _{q}\right)
e^{iq\left( R_{j}-R_{l}\right) }  \label{Veffective}
\end{equation}
\begin{equation}
t_{jl}^\ast=\tau t_{jl} \qquad X_{jl}^\ast=\tau X_{jl}\qquad
P_{jl}^\ast=\tau^4P_{jl}\qquad J_{jl}^\ast=J_{jl} \label{tXP}
\end{equation}
\begin{equation}
\tau=\exp \left\{ - \frac{1}{L}\sum_{q}\left({{g_q}\over{
\hbar\Omega_{q}}}\right)^2 \delta _{q}^{2}\left[ 1-\cos \left(
qa\right) \right] e^{-2\alpha _{q}}\right\}\; ;  \label{tau}
\end{equation}

(iii) the eigenstates and eigenvalues of the polaronic Hamiltonian
are determined by using an exact diagonalization method proposed
in Ref.\onlinecite{canio};

(iv) the optimal values of the displacement and squeezing
parameters are determined variationally by minimizing the total
energy of the electron-phonon system.

A detailed derivation of Eqs.$\,$(\ref{epsiloneff}-\ref{tau}) is
given in Ref.\onlinecite{mazako}. We see that all the parameters,
but $J$, are renormalized to some extent by the phonons. The one-
and two-particle hopping amplitudes (respectively $t_{jl}^\ast,
X_{jl}^\ast$ and $P_{jl}^\ast$) depend on the  hopping reduction
factor $\tau $, whose value, once the optimal $\delta _q$ and
$\alpha _q$ are given, follows from Eq.(\ref{tau}). The charge
mobility is thus controlled simultaneously, and competitively, by
the displacement and the squeezing parameters.

In the following, all the bare intersite coupling constants will
be assumed non-vanishing only between nearest-neighbor sites. We
stress that, even under this assumption, the effective intersite
density-density interaction $V^*_{jl}$ (Eq.(9)) acquires a
phonon-dependent long-range contribution generally extending
beyond nearest-neighbor sites. If dispersionless phonons of
frequency $\Omega$ are assumed, then this phonon-induced
long-range part of the Coulomb interaction remains non-vanishing
only if the $q$-dependence of the displacement parameters is
explicitly accounted for. This interaction was first discussed
within LFA, for dispersive phonons, in ref.\onlinecite{alexran},
but with no attempt of estimating its value. A more recent
variational study \cite{lamagna}, employing a non-factorized
fermion-boson wavefunction in the context of a simple Hubbard
electronic Hamiltonian (i.e., such that
$X_{ij}=V_{ij}=J_{ij}=P_{ij}=0$), also found an analogous
effective interaction.

\section{Results}

\noindent The general procedure illustrated  in Section 2 has been
applied to the case of a four-site chain with dispersionless
phonons of frequency $\hbar\Omega=0.1$ eV, a value appropriate to
cuprates and manganites. The eigenvalues and the eigenvectors of
the effective Hamiltonian have been exactly calculated using the
method outlined in Ref.\onlinecite{canio}. The knowledge of all
the eigenvalues for given $a$ and $\Gamma$ allows to identify the
ground state by the independent and simultaneous optimization of
the wavevector-dependent variational parameters defining, for each
phonon mode $q$, the degree of displacement ($\delta_q$) and
squeezing ($\alpha_q$). Once the optimal values of the parameters
are identified, the full eigenvalue spectrum can be obtained. The
value of $\Gamma$ might also have been obtained variationally, as
in Ref.\onlinecite{acquarone,spa92}, but due to the increased
complexity of the numerical analysis we have investigated the
phase diagram only for three representative values of $\Gamma$,
equal to 0.8, 1, 1.2 \AA$^{-1}$. We find that $\Gamma^{-1}$
roughly scales as $a$ in determining the main features of the
phase diagram. The differences in the phase diagram for different
$\Gamma$ values at given $N$ are only quantitative, and correspond
to slight variations of the border lines between the different
phases. In particular, we have found that for increasing $N$ the
effects induced by variations of $\Gamma$ tend to become less and
less significative. Therefore, we will limit ourselves to the
discussion of the results obtained in the case
$\Gamma=1\,$\AA$^{-1}$. A study of the half-filled case for fixed
$a$ and varying $\Gamma$ has already been published\cite{noi99}.

Given the operators $A_i$ and $B_j$, defined at sites $i$ and $j$,
we evaluate the correlation function (CF) $\langle A_i \, B_j
\rangle$ in the ground state $|G\rangle$ of the effective
Hamiltonian $H^*$ according to:
\begin{equation}
\langle A_i \, B_j\rangle = \langle G|\langle 0_{ph}|e^Se^D A_i \,
B_j e^{-D}e^{-S}|0_{ph}\rangle|G\rangle \quad . \label {CFdef}
\end{equation}
The nature of the ground state is investigated for different
fillings by evaluating the dependence on the lattice constant of
the spin and charge CFs
\begin{eqnarray}
\chi^{s}_{n} & = & \frac{1}{L} \sum_{i} \langle\mathbf{S}_i \cdot
\mathbf{S}_{i+1}\rangle
\\
\chi^{s}_{nn} & = & \frac{1}{L} \sum_{i}  \langle \mathbf{S}_i
\cdot \mathbf{S}_{i+2}\rangle
\\
\chi^{c}_{n} & = &\frac{1}{L} \sum_{i}  \langle n_i~n_{i+1}\rangle
\\
\chi^{c}_{nn} & = & \frac{1}{L} \sum_{i} \langle
n_i~n_{i+2}\rangle \quad ,
\end{eqnarray}
with $\chi^{s,c}_{n}$ and $\chi^{s,c}_{nn}$ indicating the nearest
and next-nearest neighbor CFs for the spin ($s$) and charge ($c$)
channel, respectively. In addition, we have analyzed the behavior
of the on-site, nearest- and next-nearest neighbor
charge-deformation CFs, given respectively by:
\begin{eqnarray}
\Phi_{loc} & = & {{1}\over{L}}\sum_j\langle
n_j(b^{\dagger}_j+b^{~}_j)\rangle \\ \Phi_{a} & =
&{{1}\over{L}}\sum_j\langle n_j(b^{\dagger}_{j\pm a}+b^{~}_{j\pm
a})\rangle \\ \Phi_{2a} & = & {{1}\over{L}}\sum_j\langle
n_j(b^{\dagger}_{j\pm 2a}+b^{~}_{j\pm 2a})\rangle \quad .
\end{eqnarray}
It is easy to show that the above charge-deformation CFs can be
expressed as weighted sums of the charge CFs with the real-space
displacement parameters acting as weights. In the case of
dispersionless phonons their expressions are
\begin{eqnarray}
\phi_{loc} & = & {g\over{2\hbar\Omega}}\,\left( n\delta_{loc} + 2
\delta_a \chi^c_n + \delta_{2a} \chi^c_{nn}\right) \label{philoc}
\\ \phi_a & =& {g\over{2\hbar\Omega}}\,\left[
(\delta_{loc}+\delta_{2a}) \chi^c_n + \delta_{a} \chi^c_{nn} +
n\delta_{a} \right]  \label{phia}  \\ \phi_{2a} & = &
{g\over{2\hbar\Omega}}\,\left(2\delta_{a}\chi^c_n +
\delta_{loc}\chi^c_{nn} + n \delta_{2a} \right) \quad ,
\label{phi2a}
\end{eqnarray}
where $n=N/L$ is the electron density and $\delta_{loc}$,
$\delta_{a}$ and $\delta_{2a}$ are real-space displacement
parameters, associated with the coupling of an electron to a
lattice distortion, respectively on the same site, on nearest and
on next-nearest neighbor sites. Recalling that for a four-site
chain the wavevector $q$ takes the 4 values $-\pi/2a$, 0, $\pi/2a$
and $\pi/a$, and assuming $\delta_{-\pi/2a}=\delta_{\pi/2a}$ and
$\alpha_{-\pi/2a}=\alpha_{\pi/2a}$, one gets that the expressions
of $\delta_{loc}$, $\delta_{a}$ and $\delta_{2a}$ in terms of the
$q$-dependent displacement parameters are
\begin{eqnarray}
\delta_{loc} \equiv \delta_{ii} & = & {1\over 4}\,\left(\delta_0 +
2\delta_{\pi/2a} + \delta_{\pi/a} \right) \\ \delta_{a} \equiv
\delta_{i,i+1} & = & {1\over 4}\,\left(\delta_0 - \delta_{\pi/a}
\right)
\\ \delta_{2a} \equiv \delta_{i,i+2} & = & {1\over 4}\,\left(\delta_0
- 2\delta_{\pi/2a} + \delta_{\pi/a} \right)
\end{eqnarray}
(similar expressions are obtained for the real-space squeezing
parameters $\alpha_{ij}$). It is worth noting that in the case of
a single electron (i.e. $N=1$), Eqs.(\ref{philoc})-(\ref{phi2a})
give a simple relation of proportionality between the functions
$\Phi$ and the parameters $\delta$ evaluated at the same site
distance.

In the numerical analysis for $N \ge 2$ we have paid special
attention to the effect on the phase diagram of the phonon-induced
charge interaction, by distinguishing the results obtained by
including both nearest ($V^\ast_{i,i\pm 1}$) and next-nearest
($V^\ast_{i,i\pm 2}$) inter-site charge interactions, from those
obtained for $V^\ast_{i,i\pm 2}=0$. The behavior of the
correlation functions and the renormalized electronic interactions
discussed in the next sections, always refers to the case of
$V^\ast_{i,i\pm 1}$ and $V^\ast_{i,i\pm 2}$ both non-vanishing,
with the parameter $\Gamma$ chosen equal to $1\,$\AA$^{-1}$. We
have also verified that at any filling the parameters $X^\ast$,
$P^\ast$ and $J^\ast=J$ are very small and have no influence on
the phase diagram. Therefore their behavior will not be shown in
the following. We shall analyze the results for the fillings
$N=1,2,3,4$ separately.

\subsection{The case $N=1$}

\begin{figure}
\centerline{\psfig{figure=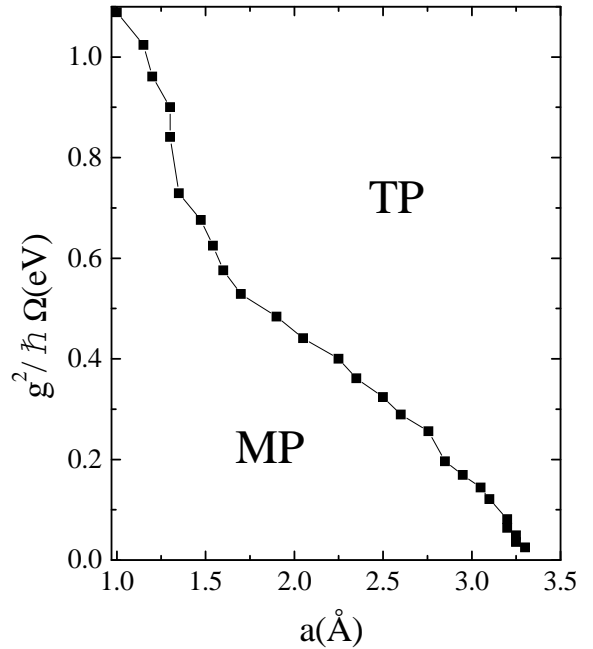,width=8cm}} \caption{Phase
diagram for $N=1$ and $\Gamma=1\,$\AA$^{-1}$. The labels MP and TP
stand for mobile polaron and trapped polaron, respectively.}
\label{fig1}
\end{figure}

We treat this case mainly to test the reliability of our approach
as compared with other methods developed in the literature
\cite{hollit,lamagna,yonemitsu,cataudella,marsiglio,chunli,alexran}.
The phase diagram for $N=1$ is shown in Fig.\ref{fig1}.

The boundary separates two regions. In the lower one the polaron
has a non-vanishing hopping amplitude, with the deformation
involving, though very weakly, the first and second neighbor
sites. In the upper region $t^\ast$ vanishes together with
$\Phi_a$ and $\Phi_{2a}$, so that the polaron is trapped and the
deformation is strictly local. This picture is in
substantial agreement with the results presented in
Refs.\cite{cataudella,marsiglio}.

\subsection{$N=2$, the quarter-filling case}

\begin{figure}
\centerline{\psfig{figure=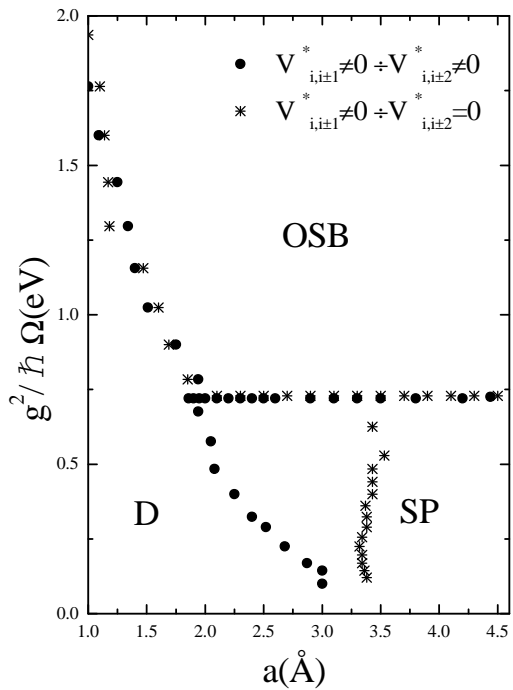,width=8cm}} \caption{Phase
diagram for $N=2$ and $ \Gamma=1\,$\AA$^{-1}$. The region labelled
D, SP and OSB are characterized by the presence of, respectively,
charge and spin disorder, separated polarons and on-site
bipolaron.} \label{fig3}
\end{figure}

The phase diagram concerning the quarter-filling case ($N=2)$ is
plotted in Fig.\ref{fig3} for $\Gamma=1\,$\AA$^{-1}$ comparing the
cases with and without $V_{i,i\pm 2}^{\ast }$. We can distinguish
three main regions: a disordered (D) phase without magnetization,
a phase in which the two electrons behave as separated polarons
(SP) in a superposition of configurations of the $1-0-1-0$ type,
and a phase characterized by the formation of an on-site bipolaron
(OSB), i.e. a superposition of configurations of the $2-0-0-0$
type.

We see that in this case the Coulomb interaction between
next-nearest neighbor sites $V^\ast_{i,i\pm 2}$ affects the phase
diagram rather strongly. Indeed, while there is no significant
effect on the boundary of the OSB region, the inclusion of the
effect of $V^\ast_{i,i\pm 2}$ leads to a considerable enlargement
of the D region at the expenses of the SP region, additionally
changing the D-SP transition from sharp to smooth.

\begin{figure}
\centerline{\psfig{figure=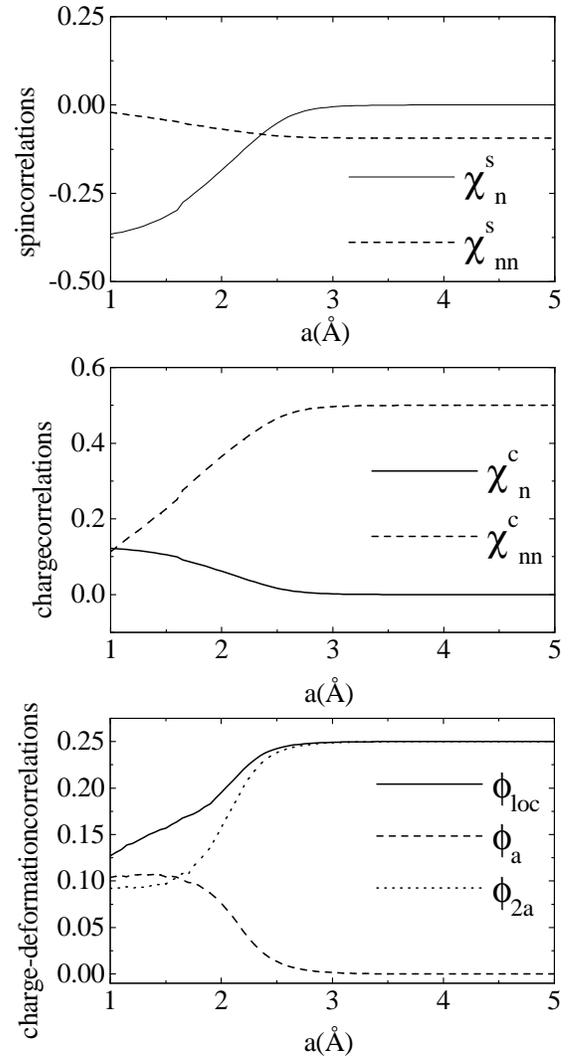,width=9cm}} \caption{Spin
(top panel), charge (middle panel) and charge-deformation (bottom
panel) correlation functions for $N=2$ at $g^2/\hbar\Omega=0.1$
eV.} \label{fig4}
\end{figure}

Let us discuss in more detail the D-SP crossover for small polaron
binding energy, choosing $g^2/\hbar\Omega =0.1\,$eV. As one can
see from Fig.\ref{fig4} (top panel), the behavior of the spin CFs
reveals a competition between different spin configurations in the
ground state, without a clear predominance of one of them. In
particular $\chi^{s}_{nn}$ is almost zero while $\chi^{s}_{n}$ is
negative in the D region revealing a tendency towards the AF
configuration. On the other hand, the charge CFs in the middle
panel of Fig.\ref{fig4} show that above a critical value of $a$
(in the case considered in the figure equal approximately to
2.8$\,$\AA), the charge distribution changes, giving rise to a
pattern $1-0-1-0$ of alternating singly-occupied and empty sites
which is characteristics of what we have called the SP region.

On the contrary, in the D region, $\chi^{c}_{n}$ and
$\chi^{c}_{nn}$ are smoothly changing with $a$, revealing the
presence of a fluctuating regime with disordered charge
distribution. Let us stress that in the SP configuration there is
no predominant character of either  FM or AF correlations, since
the slightly negative values of $\chi^{s}_{nn}$ cannot be
considered as a sign of the onset of an antiferromagnetic order.
The displacement parameters and the charge-displacement
correlation functions reveal additional interesting details of the
transition.

Actually, the bottom panel of Fig.$\,$\ref{fig4} shows that there
is a smooth, yet pronounced, change in the phononic regime for $a$
approximately greater than $2\,$\AA, thus well inside the D
region. Indeed, for small values of $a$, all the
charge-deformation CFs are non-negligible and comparable. Then,
when $a$ is increased above 2$\,$\AA $\,$, $\Phi_{loc}$ and
$\Phi_{2a}$ grow, both converging to the value 0.25, while
$\Phi_a$ decreases becoming negligible above $a=3\,$\AA. The
resulting smooth crossover from the D to the SP phase shows that
already in the D phase there are precursory effects of the charge
ordering of the $1-0-1-0$ type that characterizes the SP phase,
stable at higher values of the intersite distance.

\begin{figure}
\centerline{\psfig{figure=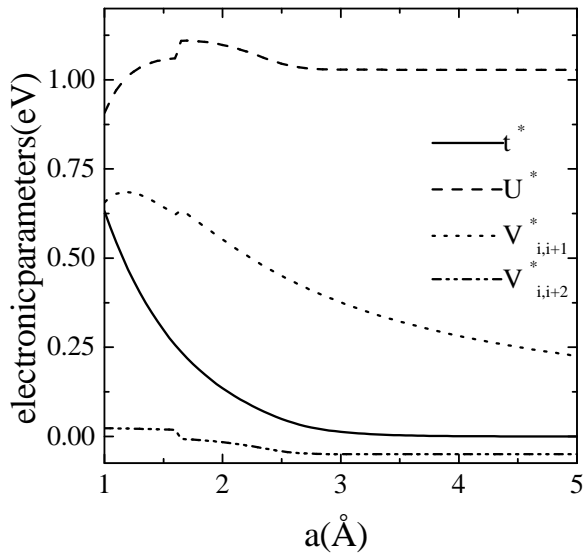,width=8cm}}
\caption{Renormalized electronic interaction parameters for $N=2$
and $g^2/\hbar\Omega=0.1\,$eV.} \label{fig7}
\end{figure}

The corresponding behavior of the effective two-body parameters is
shown in Fig.\ref{fig7}. In particular, it is interesting to
notice that a signature of the two phononic regimes in the D phase
discussed above is clearly visible in the behavior of the
next-nearest neighbor interaction $V^\ast_{i,i\pm 2}$, which
already at $a\simeq 1.7\,$\AA$\,$ turns from repulsive to
attractive, while both $U^\ast$ and $V^\ast_{i,i\pm 1}$ keep being
repulsive. This  situation clearly promotes  the SP type of charge
order, which, however, can be stable only when the effective
hopping amplitude $t^\ast$ becomes negligible. It is worth
pointing out that the D and SP states are very close in energy, so
that even a small interaction like $V_{i,i\pm 2}$ can
significantly influence the position of the phase boundary.

\begin{figure}
\centerline{\psfig{figure=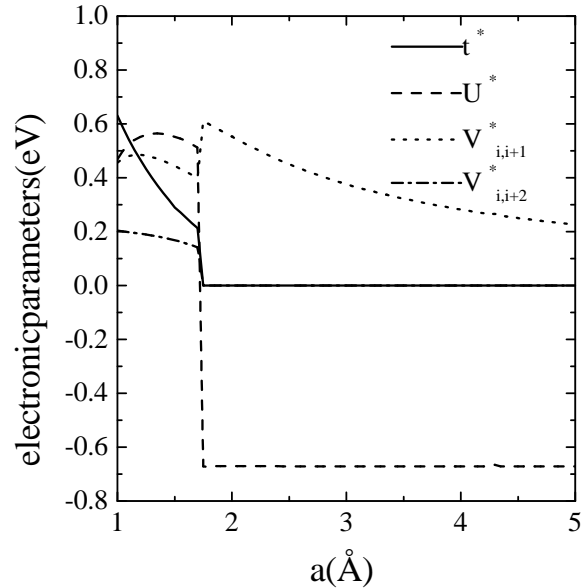,width=8cm}}
\caption{Renormalized electronic interaction parameters for $N=2$
at $g^2/\hbar\Omega=0.9$ eV} \label{fig9}
\end{figure}

Considering values of the polaron binding energy $g^2/\hbar\Omega$
greater than approximately 0.7$\,$eV, we have that the increase of
$a$ induces a transition from the disordered phase to a new charge
ordered phase, in which the local effective interaction $U^\ast$
becomes strongly attractive, thus leading to the formation of an
on-site localized bipolaron (OSB). The nature of the OSB region and
the driving mechanism leading to
its formation are clarified by the behaviour of the
renormalized electronic parameters. As one can see from
Fig.\ref{fig9}, in the OSB phase the on-site Coulomb interaction
is so strongly renormalized that it becomes attractive, while at
the same time the kinetic energy vanishes. This behavior of
$U^\ast$ and $t^\ast$ leads to the formation of
a localized on-site bipolaron.

We have also found that in the OSB phase both $\chi^{s}_{n}$ and
$\chi^{s}_{nn}$ vanish for any $a$, so that the phase is
characterized by the absence of both spin correlations and local
moments. In addition, the charge correlation functions
$\chi^{c}_{n}$ and $\chi^{c}_{nn}$ are also identically vanishing,
while $\Phi_{loc}$ is the only charge-deformation CF which is
different from zero. These results clearly indicate that in the
OSB phase the charge, together with its accompanying deformation,
is completely localized on one of the chain sites, in a
superposition of configurations of the $2-0-0-0$ type. This has
also been verified by checking that in the ground state
wavefunction all the components of the $1-1-0-0$ type have
negligibly small amplitudes.

It is interesting to compare the present results with other
studies on the behavior of two fermions coupled to lattice
deformations. The main issue discussed in the literature is to
clarify under which conditions a bound state of two polarons, i.e.
a bipolaron, either on-site or inter-site, is stable, and possibly
itinerant. In the large $g^2/\hbar\Omega$ region, dominated by the
negative value of $U^\ast$, our result of an essentially localized
on-site bipolaron agrees with those in the
literature\cite{marsiglio}. We have also verified that, even
imposing the vanishing of the bare inter-site charge repulsion
$V_{ij}$, the OSB region is not changed.

In the low $g^2/\hbar\Omega$ region, we do not find mobile
inter-site bipolarons, at variance with other
studies\cite{lamagna,yonemitsu,bonca,marsiglio,wellein,chunli,alexandrov00}.
In assessing the stability of either type of bipolaron, it is very
important to take into account the bare contribution $V_{i,i\pm
1}$ to the effective inter-site charge interaction $V^\ast_{ij}$.
In the papers in which a stable inter-site bipolaron is found,
this term, that is repulsive and usually rather large (a sizeable
fraction of $U$), is however
neglected\cite{lamagna,bonca,marsiglio,wellein,chunli}, or
explicitly assumed to be renormalized to an overall attractive
value\cite{yonemitsu}. Another recent paper\cite{alexandrov00}
emphasizes the physical relevance of the inter-site bipolarons.
Within its LFA treatment of the long-range Fr\"{o}hlich
electron-phonon interaction with dispersive phonons, the bare
$V_{ij}$ is included, but the equivalent of our $V^\ast_{ij}$ is
estimated, on a purely phenomenological basis, to be very small,
even between nearest neighbors.

Our main conclusion on the formation of a bipolaron and, in
particular, on its size, is that, when the complete set of bare
two-particle interactions is taken into account, for $N=2$ only
on-site bipolarons are possible, this taking place in the range of
parameters corresponding to the OSB region of Fig.\ref{fig3}. When
stable, the OSB has a localized character, due to the simultaneous
vanishing of the effective single-particle hopping $t^\ast$ and
the effective pair hopping amplitude $P^\ast$. Hence, our results
do not support the presence of intersite mobile bipolarons for
realistic values of the interactions, even if on the other hand
they cannot definitely rule them out in infinite lattices.

Besides the limitation of dealing with a cluster, one might also
question the approach that we adopt here to evaluate $V_{ij}$
\cite{acquarone}, yielding rather high values of the bare
intersite repulsion that, for physically reasonable values of the
phonon frequency $\Omega$ and of the coupling strength $g$, can be
only partially compensated by the phonon renormalization. However,
even if the absolute magnitudes of the various interactions we
evaluate are model-dependent, their ratios should be of more
general applicability. Indeed, it is known that in several classes
of physical systems the nearest-neighbor Coulomb interaction is a
significative fraction of the on-site one. This is, for instance,
the case of the high-$T_c$ superconducting copper oxides, where
theoretical estimates of $V_{i,i+1}$ \cite{becca,feiner} give a
repulsive value of the order of 0.2$\,$eV.

\subsection{$N=3$, one-hole case}

\begin{figure}
\centerline{\psfig{figure=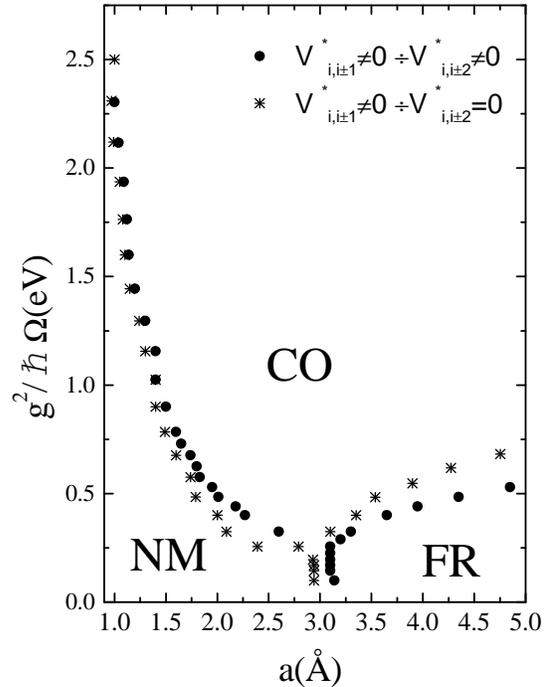,width=8cm}} \caption{Phase
diagram for $N=3$ and $\Gamma=1\,$\AA$^{-1}$. The regions labelled
NM, CO and FR are characterized, respectively, by vanishing
magnetization, charge order and frustrated magnetic order.}
\label{fig11}
\end{figure}

The phase diagram for $N=3$ is shown in  Fig.\ref{fig11}. We start
by discussing the phases that are stable at low values of the
polaron binding energy, looking at the behaviour of the
correlation functions for $g^2/\hbar\Omega=0.1\,$eV.

\begin{figure}
\centerline{\psfig{figure=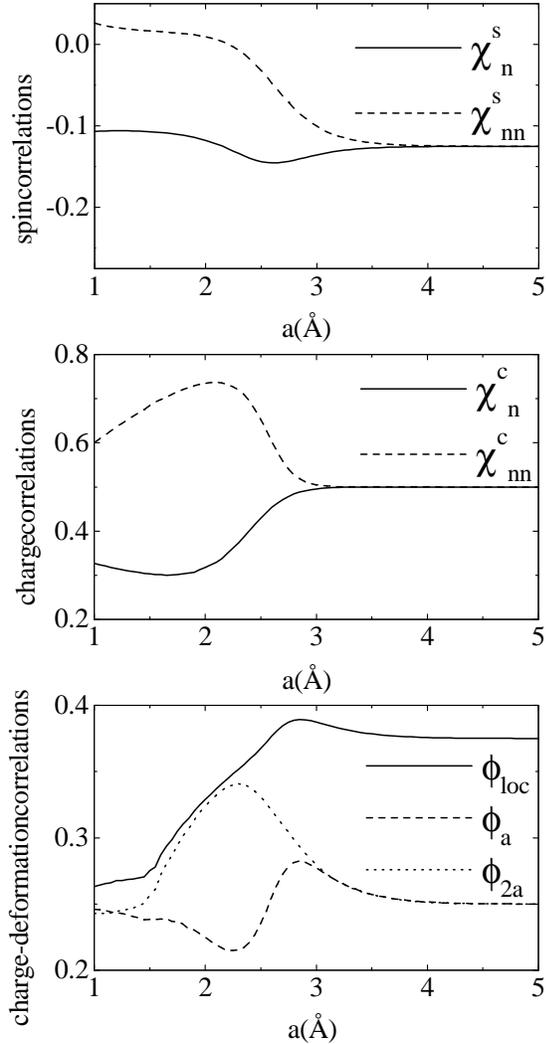,width=9cm}} \caption{Spin
(top panel), charge (middle panel) and charge-deformation (bottom
panel) correlation functions for $N=3$ at
$g^2/\hbar\Omega=0.1\,$eV.} \label{fig12}
\end{figure}

As one can see from the upper panel of Fig.\ref{fig12}, the
behavior of $\chi^{s}_{n}$ shows little changes as $a$ is varied
and it is always negative. On the other hand, $\chi^{s}_{nn}$ is
positive and close to zero for $a$ smaller than $\simeq 2\,{\rm
\AA}$ and then it changes sign as $a$ is increased, going to a
saturation value of about $-0.125$. As far as the charge channel
is concerned, we see from the middle panel of Fig.\ref{fig12} that
$\chi^{c}_{n}$ exhibits slight variations with $a$, with a smooth
increase concentrated in a range going approximately from
2$\,$\AA$\;$ to 3$\,$\AA. On the other hand, up to $a\approx
3\,$\AA$\,$ $\chi^{c}_{nn}$ is larger than $\chi^{c}_{n}$, first
growing (up to $a\approx 2\,$\AA) and then smoothly decreasing
towards the saturation value $0.5$ that both $\chi^{c}_{n}$ and
$\chi^{c}_{nn}$ reach for $a>3$\AA.

As far as the charge-deformation CFs are concerned, we see from
the bottom panel of Fig.\ref{fig12} that $\Phi_{loc}$ is the
largest at any $a$, with $\Phi_{2a}$ manifesting a significative
increase exactly in the range of intermediate values of $a$ in
which $\chi^{c}_{nn}$ reach its maximum. In addition $\Phi_a$
turns out to be the smallest charge-deformation CF at any $a$,
becoming equal to $\Phi_{2a}$ only above $a\approx 3\,$\AA.

Combining the results for the CFs we conclude that for $a$ lower
than $\approx 2\,$\AA$\,$ in the non-magnetic (NM) region, the
charge is distributed with fluctuating configurations of almost
empty and almost doubly occupied next-nearest neighbor sites. This
causes both $\chi^{c}_{nn}$ and $\chi^{s}_{nn}$ to be negligible.
Then, as $a$ is increased beyond $2\,$\AA$\,$ up to $a\approx
3\,$\AA, there is, consistently with the behavior of $\Phi_{2a}$,
a crossover region which can be seen as a precursor of the charge
order of the $2-0-1-0$ type, which characterizes the ground state
of the system at higher values of $g^2/\hbar\Omega$. Finally, the
charge distribution tends to become uniform as the system, for $a
>3$ \AA, enters the region that we label as ``frustrated
magnetism'' (FR). Indeed, there $\chi^{s}_{n}$ and $\chi^{s}_{nn}$
converge to the same unsaturated negative value, which expresses a
tendency of the spins to align antiferromagnetically both on
nearest- and next-nearest neighbor sites. The FR zone is also
characterized by a reduction in the extent of the deformation
around each occupied site, as indicated by $\Phi_{2a}$ becoming
the smallest charge-deformation CF.

The renormalized electronic parameters (not shown) vary smoothly
with $a$. They never change sign, with the exception of the
long-range charge interaction which, always very small, changes
from positive to negative near the NM-FR boundary.

\begin{figure}
\centerline{\psfig{figure=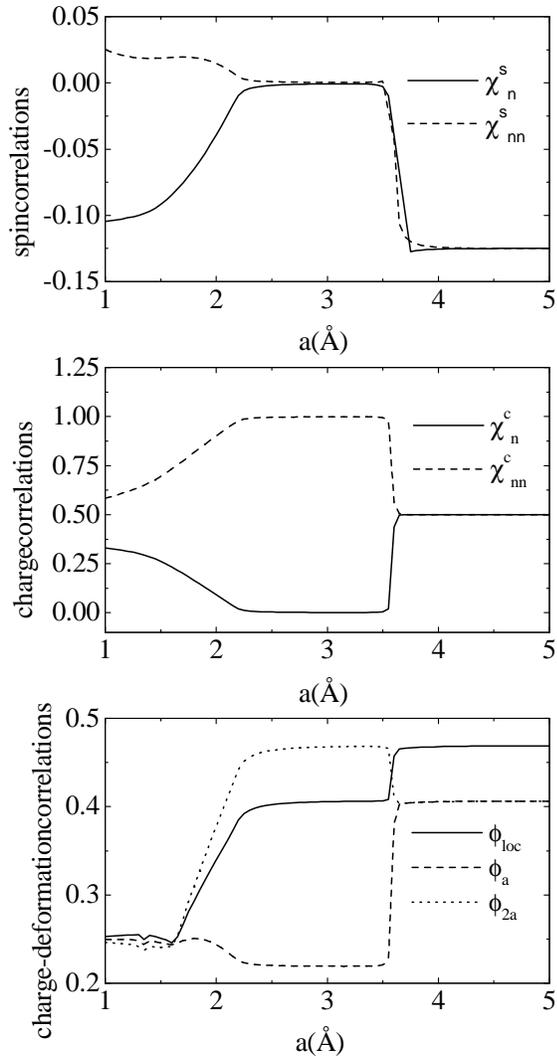,width=9cm}} \caption{Spin
(top panel), charge (middle panel) and charge-deformation (bottom
panel) correlation functions for $N=3$ at
$g^2/\hbar\Omega=0.4\,$eV.} \label{fig15}
\end{figure}

Let us now consider larger values of the polaron binding energy,
investigating in particular that part of the phase diagram,
corresponding to values of $g^2/\hbar\Omega$ between approximately
0.25$\,$\AA $\,$ and 0.7$\,$\AA, where a new charge ordered (CO)
phase develops between the NM and the FR regions for intermediate
values of $a$. This will be done with a special attention to the
character of the CO phase, by analyzing the behavior of the CFs at
the representative value $g^2/\hbar\Omega =0.4\,$eV. As one can
see from the upper panel of Fig.\ref{fig15}, for values of $a$
falling in a range going approximately from 2.2$\,$\AA$\,$ to
3.7$\,$\AA$\,$ the spin CF $\chi^{s}_{n}$ and $\chi^{s}_{nn}$ are
zero, indicating the absence of local moments and magnetic
correlations in the ground state. On the other hand, by looking at
the behaviour of $\chi^{c}_{n}$ and $\chi^{c}_{nn}$ in the same
range of values of $a$, we can see that charge correlations are
vanishing on nearest-neighbor sites and tend to the maximum value
compatible with the filling on next-nearest neighbor sites. This
is a clear indication of the development of an ordered
configuration where the charge is fully distributed on
next-nearest neighbor sites.

The same general trend can be inferred from the charge-deformation
CFs, shown in the lower panel of Fig.\ref{fig15}. We see that the
correlations between the charge and the deformations in the CO
region become the strongest between second neighbor sites, with
the first neighbor one being the smallest. This can be interpreted
as a further indication that the ground state is given by a
superposition of the charge configurations $1-0-2-0$ and
$2-0-1-0$, together with the symmetrically related $0-1-0-2$ and
$0-2-0-1$. This degeneracy of the ground state, related to the
fact that we cannot have a really broken-symmetry state in
finite-size systems, is also the reason why in the CO region
$\chi^{c}_{nn}$ is equal to 1, rather than being equal to 2. The
whole picture is also consistent with the behavior of the
renormalized electronic parameters,
featuring a non-negligible negative value of $V^\ast_{i,i\pm 2}$
both in the CO and in the FR region.

The filling $N=3$ can be viewed as the case of one hole in an
half-filled configuration. There are two different predictions on
the related ground state in the limit of large $U^\ast/t^\ast$.
According to Nagaoka's theorem, a ferromagnetic (FM) state should
set in when $U\to\infty$ with $t$ remaining finite, while
according to Refs.\cite{deeg,yonemitsu,wellein,fehske97} the hole
would be a polaron in an antiferromagnetic (AF) background, where
its phonon-induced localization would take place at a lower value
of $g^2/\hbar\Omega$ due to concurrent effect of the ordered spin
arrangement. In our approach we find that for $N=3$ none of the
two magnetic ground states mentioned above is stable. On the
contrary, in what we have called the frustrated region no magnetic
effect prevails, this region being characterized by FM and AF
correlations of comparable amplitude. While one can only speculate
on the fact that we do not find an AF ground state, guessing that
it is a finite-size effect, on the other hand the reason why we
find no FM order is that the conditions under which Nagaoka's
theorem holds are never satisfied. Indeed there is no parameter
regime where the phononic renormalization yields $U^\ast \gg
t^\ast$, with $t^\ast$ remaining at the same time finite.

\subsection{ $N=4$, the half-filled case}

\begin{figure}
\centerline{\psfig{figure=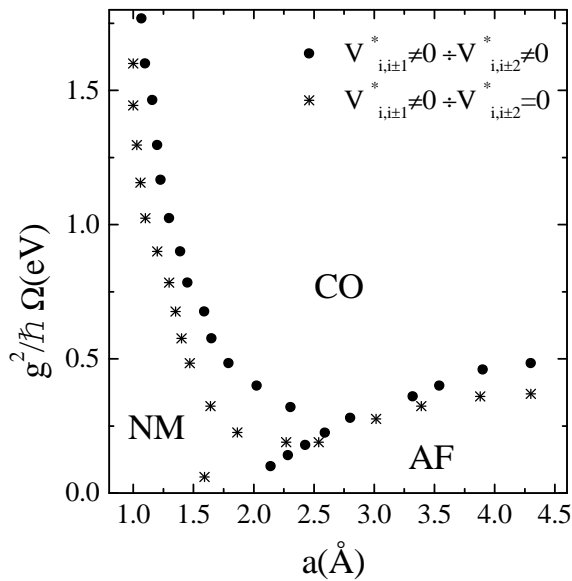,width=8cm}} \caption{Phase
diagram for $N=4$ and $\Gamma=1\,$\AA$^{-1}$. The regions labelled
NM, CO and AF are characterized, respectively, by vanishing
magnetization, charge order and antiferromagnetism.} \label{fig17}
\end{figure}

In Fig.\ref{fig17} we report the phase diagram at half-filling for
$\Gamma=1\,$\AA$^{-1}$, both with and without the phonon-induced
next-nearest neighbor interaction. For intermediate to large $a$
we see that an antiferromagnetic (AF) phase is stable for small
$g^2/\hbar\Omega$ while a charge ordered state (CO) of the
$2-0-2-0$ type is stable for large $g^2/\hbar\Omega$.  For small
$a$ and irrespective of $g^2/\hbar\Omega$, a non-magnetic (NM)
ground state characterized by the absence of any spin correlations
is found to be stable, wedging in between the AF and CO regions at
small $g^2/\hbar\Omega$ up to $a\sim 2.5\,$\AA.

We can also see that taking into account the effect of $V_{i,i\pm
2}^{\ast }$ leads to quantitatively significant modifications of
the phase diagram. Indeed, even if the negative value of
$V^\ast_{i,i\pm 2}$ is rather small for low $g^2/\hbar\Omega$
values, it causes an appreciable expansion of the NM region
towards higher $g^2/\hbar\Omega$ and $a$ values, with a
corresponding reduction of the CO and AF regions, due to the fact
that for increasing $g^2/\hbar\Omega$ (or $a$) $V^\ast_{i,i\pm 2}$
becomes negative, and thus opposite in sign to $V^\ast_{i,i\pm
1}$.

\begin{figure}
\centerline{\psfig{figure=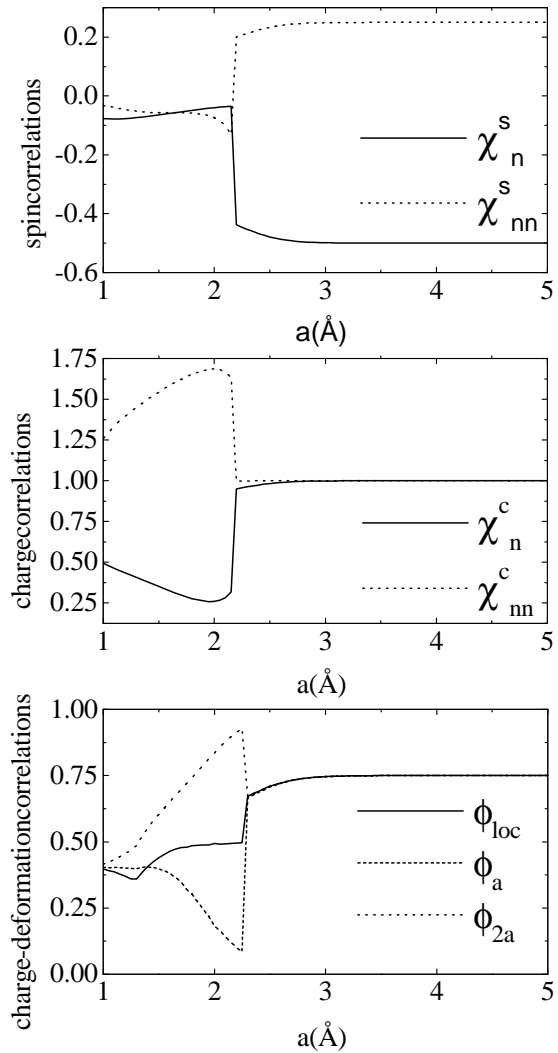,width=9cm}} \caption{Spin
(top panel), charge (middle panel) and charge-deformation (bottom
panel) correlation functions for $N=4$ at
$g^2/\hbar\Omega=0.1\,$eV.} \label{fig18}
\end{figure}

Let us now discuss how the system responds as the value of $a$ is
increased, for $g^2/\hbar\Omega =0.1\,$eV. As shown in
Fig.\ref{fig17}, the system passes from the NM to the AF regime.
The position and the nature of this transition can be easily
identified by looking at the behavior of the correlation functions
plotted in Fig.\ref{fig18}.

For $a$ lower than $\simeq 2.2\,{\rm \AA}$, $\chi^{s}_n$ and
$\chi^{s}_{nn}$ are close to zero and almost coinciding, which
corresponds to a regime without significative magnetic
correlations. On the other hand, for higher values of $a$ one has
$\chi^{s}_n\approx -0.5$, which corresponds to an antiparallel
configuration of spins on first neighbors, and
$\chi^{s}_{nn}\approx 1/4$ which indicates a configuration of
parallel spins on next-nearest neighbors. This behaviour of
$\chi^{s}_n$ and $\chi^{s}_{nn}$ clearly shows that the ground
state has antiferromagnetic correlations. The middle panel of the
Fig.\ref{fig18} gives indications on the charge distribution. In
the NM region we can see that even though the site occupation is
still fluctuating because of the itinerancy of the carriers,
nonetheless there is some precursor of the CO phase, as a
consequence of the fact that the values of $\chi^{c}_{nn}$ are
considerably higher than those of $\chi^{c}_{n}$. Conversely, in
the AF region $\chi^{c}_{n}$ and $\chi^{c}_{nn}$ become constant
and equal to one, thus indicating that there is one electron per
site on the average.

The same qualitative trend is found in the charge-deformation CFs,
shown in the bottom panel of Fig.\ref{fig18}. At small $a$ the
charge-related deformation appreciably affects also   first and
second neighbor sites. Then, as the NM-AF boundary is approached,
$\Phi_{loc}$ tends to stay constant, while $\Phi_{2a}$ grows
considerably and $\Phi_a$ becomes small, resulting in a behaviour
that can be considered as a precursor of the CO state. When the AF
state sets in, all three charge-deformation CFs come to coincide
at a rather large value, indicating that, in a regular pattern of
site occupancy, all orders of neighbors are equally affected. In
terms of the size of the polarons, one can say that for $N=4$ they
tend to show an extended character, i.e. the associated
deformation develops beyond the site where the charge is located.

The behavior of the renormalized electronic parameters (not shown)
is such that in the AF region $t^{\ast}$ is vanishingly small,
driving the system in a regime of electron localization where the
mechanism of superexchange dominates. Conversely, in the NM
region, the hopping term is only weakly renormalized, because the
localization effect due to the displacement is counteracted by the
squeezing, so that the electrons are rather mobile. Moreover,
$U^\ast$ and $V_{i,i\pm 1}^{\ast}$ are comparable in magnitude,
thus preventing the formation of local pairs as well as of long
range ordering. In terms of an equivalent $t-J$ model, the NM
ground state does not show any well-defined spin or charge order
since $t^{\ast}$ remains large enough to have two concurrent
effects: on one side it opposes a complete charge
disproportionation, and on the other side it gives rise to a
kinetic exchange proportional to $2 t^{\ast 2}/(U^{\ast }-V_{i\pm
1}^{\ast })$ strong enough to allow for some magnetic coupling
effects. As in the previous cases, the energy difference between
the NM and the AF states is small enough for the tiny negative
value of $V^\ast_{i,i\pm 2}$ to have a sizeable effect on the
position of the phase boundaries.

\begin{figure}
\centerline{\psfig{figure=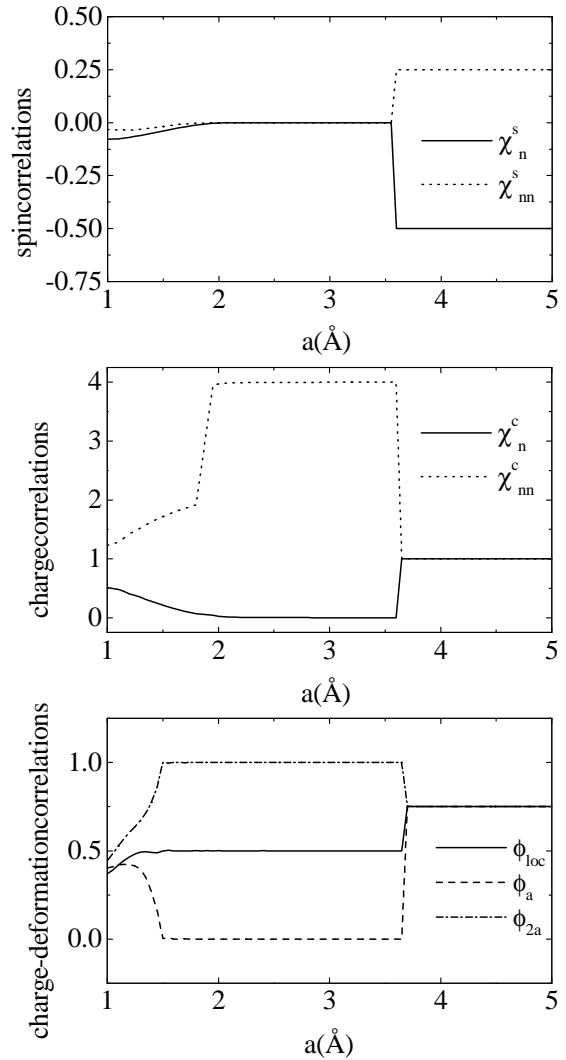,width=9cm}} \caption{Spin
(top panel), charge (middle panel) and charge-deformation (bottom
panel) correlation functions for $N=4$ at
$g^2/\hbar\Omega=0.4\,$eV.} \label{fig21}
\end{figure}

When $g^2/\hbar\Omega=0.4\,$eV, a reentrant behavior can be
deduced from the phase diagram in Fig.\ref{fig17} as $a$ is
increased. This is made evident by the behavior of the correlation
functions in Fig.\ref{fig21}. For $a$ between approximately
2$\,$\AA$\,$ and 3.6$\,$\AA, $\chi^{c}_{n}$ is equal to zero
(middle panel), while $\chi^{c}_{nn}$ assumes the largest value
($\sim 4$), clearly indicating a CO configuration of alternating
empty and doubly occupied sites. As expected, in this same range
of $a$ the spin correlation functions (upper panel) are zero since
configurations with zero and two electrons on each site prevent
the formation of magnetic moments. As far as the
charge-deformation CFs are concerned, we see that in the CDW
region the largest values are taken by $\Phi_{2a}$, with the
nearest-neighbor CF $\Phi_{a}$ remaining always very small, and
the on-site CF $\Phi_{loc}$ taking values which are intermediate
between $\Phi_{2a}$ and $\Phi_{loc}$. This can clearly be
considered as a further evidence of the formation of a CO state of
the $2-0-2-0$ type. When $a$ lies out of this intermediate range,
the correlation functions behave essentially as in the case
$g^2/\hbar\Omega=0.1\,$eV.

The renormalized electronic parameters (not shown) fully support
this picture. In the CDW region the electron mobility is equal to
zero, $V^\ast_{i,i\pm2}$ is attractive and $V^\ast_{i,i\pm 1}$ is
repulsive and comparable in magnitude to $U^\ast$. This evidently
tends to promote the configurations with alternating empty and
doubly occupied sites. To the extent that the effects of including
$V^\ast_{ij}$ can be neglected, our results qualitatively agree
with other studies \cite{trapper}.

We end this Section by noting that the discontinuities in the
correlation functions and the renormalized parameters seen in the
above figures, should not be interpreted as a signature of a real
first-order localization transition. They are rather a consequence
of a drawback affecting all variational methods, which, as it is
well-known, can only provide an upper bound to the true ground
state energy. This leads to a violation of the exact results
demonstrated by L\"{o}wen \cite{exact}, according to which the
self-trapping transition generated by an interacting term of the
Holstein type is an analytical crossover for any value of the
electron-phonon coupling, and not an abrupt (non-analytical) phase
transition.

\section{Conclusions}

The method that we have used to obtain an effective polaronic
Hamiltonian from the generalized Holstein-Hubbard Hamiltonian,
based on the use of the displacement and squeezing
transformations, while allowing technically simple calculations,
yet permits to grasp some basic physical effects, such as, for
instance, the phonon-induced renormalization of the electronic
interaction parameters. In this respect, this approach might be of
interest, for instance, in the theoretical analysis of high-$T_c$
superconductivity, where it can provide clues to select among
electron-phonon models or purely electronic models. If, in the
comparison between theory and experiments, one refers to
phonon-renormalized, and not bare, interactions, then even
phenomena such as the isotope effect could find an explanation
within the framework of a model with only electronic degrees of
freedom. An improvement of the approach followed here may be
obtained by suitably dealing with two limitations, that is, (i)
the factorization of the fermion-boson wave function into separate
fermionic and bosonic factors, and (ii) the use of a squeezed
phonon wavefunction (whose intrinsic limitations have been
discussed in Ref.\onlinecite{borghi}) for the elimination of the
phononic degrees of freedom. Further studies in this direction are
planned for the next future.

The results we have presented support our claim that, in working
out an effective Hamiltonian through displacement and squeezing
transformations, it is important to take into account the momentum
dependence of the displacement and squeezing parameters, to avoid
losing a significant part of the physical content of the model. We
have indeed shown that the effective phonon-induced inter-site
charge interaction that one gets\cite{mazako,alexran} by going
beyond LFA, should be carefully taken into account, as it has
appreciable effects on the phase diagram, particularly for $N=4$
and $N=2$. This latter case is particularly significant, because
theoretical studies\cite{becca,thstripes} of the occurrence of
doping-induced inhomogeneities in the charge distribution in both
colossal magnetoresistance manganites and superconducting
cuprates, point to the importance of the long-range charge
interaction, but have difficulties in justifying its presence in
metallic systems where it should vanish beyond a very short
distance, due to the screening induced by the itinerant carriers.
Differently from the purely electronic $V_{ij}$, the effective
$V^\ast_{ij}$ can be seen as an independent parameter providing
the needed long-range interaction through a mechanism depending
basically on the phonons, and only weakly on the doping. In real
systems such as the  superconducting cuprates this mechanism can
be ascribed to the Holstein-type electron-phonon coupling
generated by the ion dynamics out of the conducting layers.
Additionally, $V^\ast_{i,i\pm 2}$ can vary both in amplitude and,
most importantly, in sign, depending on the phononic state and on
the value of the lattice constant.

>From a more general point of view, our results show that there
exist, for $N=2,3$, regions of the phase diagram where the ground
state of a hole-doped correlated system with strong electron-phonon
interaction, on one side has appreciable coexisting spin and
charge correlations, while on the other side is such that the
carriers retain, to a large extent, their itinerancy. This
suggests that the debate on whether charge or spin fluctuations in
the normal state are responsible for high-temperature
superconductivity might be viewed from a different perspective,
namely, by considering that, under appropriate conditions, both
fluctuations can coexist, possibly cooperating to build the
superconductive phase.

A significant result is that, for $N\ge 2$, in the disordered
regions of the phase diagrams for low values of $a$ and
$g^2/\hbar\Omega$, we find wide subregions characterized by strong
precursory charge-ordering effects, whose amplitude can change
very rapidly with the lattice parameter. The effect is
particularly evident in the case $N=2$. This might be of interest
with respect to the recently reported\cite{bozin} microscopically
coexisting charge and structural dishomogeneities in underdoped
and optimally doped La$_{2-x}$Sr$_x$CuO$_4$ samples.

Finally, we have been able to model the effects of a varying
lattice spacing. This might be of relevance in all the cases in
which compression (or expansion) of the lattice causes sharp phase
transitions. In particular, our results show that in the range of
$a$ where a given state is stable, even if that state does not
show a fully developed symmetry-breaking order parameter (CO or
AF), still the behavior of some correlation functions may be
characterized by drastic changes as the site distance is varied.

\bigskip\bigskip

\noindent {\bf Acknowledgements}

\medskip

Interesting discussions with R. Iglesias, M.A. Gusm\~{a}o, A.
Painelli and H. Zheng are gratefully acknowledged.

\end{document}